\documentclass[conference]{IEEEtran}
\pdfoutput=1
\usepackage{cite}
\usepackage{amsmath,amssymb,amsfonts}
\usepackage{algorithmic}
\usepackage{graphicx}
\usepackage{textcomp}
\def\BibTeX{{\rm B\kern-.05em{\sc i\kern-.025em b}\kern-.08em
		T\kern-.1667em\lower.7ex\hbox{E}\kern-.125emX}}
\usepackage{booktabs}
\usepackage{xcolor}
\usepackage{epsfig}
\usepackage{epstopdf}
\usepackage{acronym}
\graphicspath{{fig/}}
\usepackage{mathtools}
\usepackage{siunitx}

\usepackage{amsthm}

\newtheorem{statement}{Statement}

\definecolor{applegreen}{rgb}{0.55, 0.71, 0.0}
\definecolor{awesome}{rgb}{1.0, 0.13, 0.32}
\definecolor{azure(colorwheel)}{rgb}{0.0, 0.5, 1.0}
\definecolor{darklavender}{rgb}{0.45, 0.31, 0.59}
\definecolor{cyan(process)}{rgb}{0.0, 0.72, 0.92}
\definecolor{brightmaroon}{rgb}{0.76, 0.13, 0.28}
\definecolor{ao(english)}{rgb}{0.0, 0.5, 0.0}
\definecolor{brightturquoise}{rgb}{0.03, 0.91, 0.87}
\definecolor{bondiblue}{rgb}{0.0, 0.58, 0.71}

\newacro{ACDD}{Alamouti with cyclic delay diversity}
\newacro{URLLC}{ultra-reliable low-latency communications}
\newacro{3GPP}{third generation partnership project}
\newacro{PHY}{physical layer}
\newacro{MIMO}{multiple-input multiple-output}
\newacro{SIMO}{single-input multiple-output}
\newacro{MISO}{multiple-input single-output}
\newacro{SISO}{single-input single-output}
\newacro{MRC}{maximum-ratio combining}
\newacro{SNR}{signal-to-noise ratio}
\newacro{CP}{cyclic prefix}
\newacro{CDD}{cyclic delay diversity}
\newacro{FSC}{frequency-selective channel}
\newacro{STC}{space-time coding}
\newacro{FFT}{fast Fourier transform}
\newacro{LMMSE}{linear minimum mean-squared error}
\newacro{FER}{frame error rate}
\newacro{OFDM}{orthogonal frequency division multiplexing}
\newacro{OCDM}{orthogonal chirp division multiplexing}
\newacro{FSC}{frequency-selective channel}
\newacro{CSI}{channel state information}
\newacro{LMMSE-PIC}{linear mininmum mean squared error with parallel iterference cancellation}
\newacro{PFE}{perfect-feedback equalizer}
\newacro{FD}{frequency domain}
\newacro{PDP}{power delay profile}
\newacro{PDF}{probability density function}
\newacro{DFT}{discrete Fourier transform}
\newacro{ICI}{inter-carrier interference}
\newacro{OTFS}{orthogonal time frequency space}
\newacro{AWGN}{additive white Gaussian noise}
\newacro{SWH}{sparse Walsh-Hadamard}
\newacro{LLR}{log-likelihood ratio}
\newacro{PMF}{probability mass function}
\newacro{CRC}{cyclic redundancy check}
\newacro{PAM}{pulse amplitude modulation}
\newacro{QAM}{quadrature amplitude modulation}
\newacro{FWHT}{fast Walsh-Hadamard transform}
\newacro{MAP}{maximum a-posteriori}
\newacro{SC}{single-carrier}
\newacro{ISI}{inter-symbol interference}
\newacro{ZP}{zero-padding}
\newacro{BCJR}{Bahl, Cocke, Jelinek, and Raviv}
\newacro{ISAC}{integrated sensing and communication}
\newacro{DFRC}{dual function radar communication}
\newacro{UE}{user equipment}
\newacro{PSD}{power spectral density}
\newacro{BS}{base station}

\usepackage{soul,color}

\usepackage{pgfplots}
\usepackage{tikz}
\usetikzlibrary{shapes,arrows}
\usetikzlibrary{positioning,calc}
\usetikzlibrary{decorations.pathreplacing,calligraphy}

\tikzset{add/.style n args={4}{
		minimum width=3mm,
		path picture={
			\draw[black] 
			(path picture bounding box.south east) -- (path picture bounding box.north west)
			(path picture bounding box.south west) -- (path picture bounding box.north east);
			\node at ($(path picture bounding box.south)+(0,0.13)$)     {\tiny #1};
			\node at ($(path picture bounding box.west)+(0.13,0)$)      {\tiny #2};
			\node at ($(path picture bounding box.north)+(0,-0.13)$)        {\tiny #3};
			\node at ($(path picture bounding box.east)+(-0.13,0)$)     {\tiny #4};
		}
	}
}

\tikzset{add2/.style n args={4}{
		minimum width=1mm,
		path picture={
			\draw[black] 
			(path picture bounding box.south) -- (path picture bounding box.north)
			(path picture bounding box.west) -- (path picture bounding box.east);
			\node at ($(path picture bounding box.south)+(0,0.13)$)     {\tiny #1};
			\node at ($(path picture bounding box.west)+(0.13,0)$)      {\tiny #2};
			\node at ($(path picture bounding box.north)+(0,-0.13)$)        {\tiny #3};
			\node at ($(path picture bounding box.east)+(-0.13,0)$)     {\tiny #4};
		}
	}
}

\usepackage{fancyhdr}

\title{A System Level Analysis for \\ Integrated Sensing and Communication}
\author{
	\IEEEauthorblockN{Roberto Bomfin\IEEEauthorrefmark{1}, Konpal Shaukat Ali\IEEEauthorrefmark{1}, Marwa Chafii\IEEEauthorrefmark{1}\IEEEauthorrefmark{2}}
	\IEEEauthorblockA{Email: \{roberto.bomfin, konpal.ali, marwa.chafii\}@nyu.edu}
	\IEEEauthorblockA{\IEEEauthorrefmark{1}Engineering Division, New York University (NYU), Abu Dhabi, UAE.}
	\IEEEauthorblockA{\IEEEauthorrefmark{2}NYU WIRELESS, NYU Tandon School of Engineering, New York, USA}
	\vspace{-1.1cm}
}

\begin{document}

	\maketitle	
	
	\thispagestyle{fancy}
	\lhead{Bomfin, R., Ali, K., and Chafii, M., ``A System Level Analysis for Integrated Sensing and Communication,'' \textit{in IEEE Wireless Communications and Networking Conference (WCNC) 2024}, Dubai,	United Arab Emirates, April 2024. (Accepted for Publication).}

	\begin{abstract}
		In this work, we provide a system level analysis of integrated sensing and communication (ISAC) systems, where a setup with a mono-static dual-functional radar communication base station is assumed.
		We derive the ISAC \ac{SNR} equation that relates communication and radar \acp{SNR} for different distances. We also derive the ISAC range equation, which can be used for sensing-assisted beamforming applications.
		Specifically, we show that increasing the frequency and bandwidth is more favorable to the radar application in terms of relative SNR and range while increasing the transmit power is more favorable to communications.
		Numerical examples reveal that if the range for communication and radar is desired to be in the same order, the ISAC system should operate in mmWave or sub-$\SI{}\THz$ bands, whereas sub-$\SI{6}\GHz$ allows scenarios where the communication range is of orders of magnitude higher than that of radar.
	\end{abstract}
	
	\section{Introduction}
	%
	\IEEEPARstart{W}{ith} recent advancements in the wireless technologies fields, both communications and radar systems have been employing antenna array technologies and have been moving towards higher frequency bands, leading to similarities of hardware architecture, channel characteristics and signal processing.
	These fundamental convergence aspects offer the opportunity for the integration of communication and sensing into the same network infrastructure.
	For these reasons, integrated sensing and communication (ISAC) has emerged as one of the key novel technologies of the future 6G radio access network \cite{LiuJSAC,ZhangISAC}, and is currently under the initial research phase in academia and industry. 
	The envisioned applications enabled by ISAC are numerous, including smart factoring, the internet of things, robotics, environmental monitoring, and vehicular communications, among others.
	
	In order to take the most advantage of integration gains due to more degrees of freedom and flexibility, the joint design of communication and sensing waveforms is a promising approach to design ISAC systems 
	\cite{LiuX,bazzipapr}
	since it offers interesting trade-offs between communication and sensing, instead of prioritizing one application.
	One approach of joint design is to optimize the waveform in the space domain, where a \ac{DFRC} \ac{BS} optimizes transmitted beams to simultaneously serve users and create directional beams towards a desired target \cite{LiuTSP,Bazzi,Chen,Thuy}.
	Another approach is to design signals that are suitable to convey data and estimate the channel parameters, where \ac{OFDM} and \ac{OTFS} waveforms have been considered \cite{Gaudio,Raviteja,Keskin}.
	Most of the previous works in this field have concentrated on proposing novel architectures and designing signal processing algorithms for ISAC.
	However, to the best of our knowledge, there has been so far no systematic system level analysis that considers both communication and radar sensing applications. 
	We emphasize that the integration of radar and communication may be challenging due to the fundamental difference between the path loss power of communication and radar over distance. In particular, the power of communication and radar drops square and fourth power of the distance, respectively. 
	However, the system parameters such as frequency, bandwidth, number of antennas and transmit power also play a role in the power budget. 
	Thus, a detailed power budget analysis should be conducted in order to understand the role of each parameter on the SNR of communication and radar regardless of the underlying signal processing or specific task.
	
	In this work, we provide a system level analysis of ISAC, where a mono-static \ac{DFRC} \ac{BS} is considered.
	In particular, we derive the ISAC SNR operating point (SOP) expression that defines a relation between communication and radar SNRs for different distances, which allows a systematic investigation of how the system parameters impact the SNR, and consequently the performance, of both applications.
	In addition, we show how to incorporate different levels of coupling between the communication and radar channels in the SNR equation, making the investigation generic.
	Another interesting aspect is the range analysis. 
	In this case, we are interested in studying the relation between the range for both communication and radar services for a desired performance metric based on the SNR.
	This analysis is particularly important for sensing-assisted beamforming where the communication user and the target should be in the same location.
	In the numerical examples section, we show that if the range of both communication and radar is desired to be in the same order, the ISAC system should operate in mmWave or sub-$\SI{}\THz$ bands, whereas sub-$\SI{6}\GHz$ allows scenarios where the communication range is of orders of magnitude higher than that of radar.
	The contributions are summarized as:
	\begin{itemize}
		\item Derivation of the SOP and range expressions for ISAC.
		\item Analysis of how system parameters impact SNR operating points and range ratio. 
		\item Providing numerical examples and showing how the derived equations can be used to design ISAC systems.
	\end{itemize}
	
	The remainder of this paper is organized as follows. 
	Section~\ref{sec:ISAC_model} introduces the system model and defines the SNR expressions for communication and radar.
	Section~\ref{sec:SNR_analysis} derives the ISAC SNR equation and gives insights on how the system parameters impact it.
	Section~\ref{sec:analaysis_alpha_beta} generalizes the ISAC SNR equation to arbitrary channel coupling levels.
	Section~\ref{sec:range_equation} derives the range equation.
	Section~\ref{sec:numerical_examples} provides numerical examples of how our analysis can be utilized in order to design ISAC systems.
	Lastly, Section~\ref{sec:conclusion} concludes the paper.
	
	{\it Notations}: $\log(\cdot)$ is defined in the base 10. For a real number $X$, $\tilde{X} = 10\log X$.
	
	\section{ISAC Model}\label{sec:ISAC_model}
	\begin{figure}[t!]
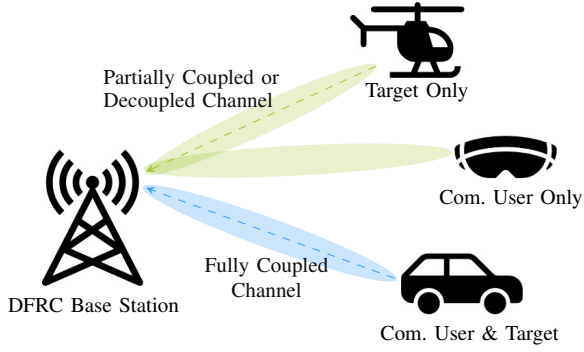

		\centering
		\include{system_model}
		\vspace{-1cm}
		\caption{ISAC Model with fully coupled, partially coupled, or decoupled channel.}
		\label{fig:system_model}
		\vspace{-0.5cm}
	\end{figure}
	\subsection{Assumptions}
	We consider an ISAC system composed of a DFRC base station (BS) with monostatic radar capability where both communication and radar signals are the same as depicted in Fig.~\ref{fig:system_model}.
	For simplicity, a single-user and single-target scenario is considered with a line-of-sight channel model and free-space path loss.
	\subsection{Channel Coupling Levels}
	In general, there are three possible channel scenarios with respect to the coupling level between the communication and radar channels, namely, {\it fully coupled}, {\it partially coupled}, and {\it decoupled}.
	The fully coupled channel occurs when the BS uses the same transmit beam for communication and radar, thereby the transmit power is the same for both services.
	We notice that the fully coupled channel admits cases where the communication user and radar target have different distances from the DFRC BS, where they are located along the same direction, i.e., azimuth angle, covered by a single transmit beam.
	The partially coupled and decoupled channels describe the situation where the user and target are located along different directions from the DFRC BS.
	For instance, the decoupled channel situation happens when there is enough angular separation between the user and the target from the DFRC BS viewpoint, where the BS is able to create orthogonal beams with distinct power towards the user and target.
	Lastly, the partially coupled channel situation happens when the transmit beams for communication and radar are partially correlated, meaning that there is a common power level that is necessarily shared by both applications. 
	The coupling level between the channels can be described by $  0\leq  \beta \leq 1$, where $\beta = 1$ and $\beta = 0$ represent fully coupled and uncoupled channel cases, respectively.
	Naturally, for $\beta < 1$, the base station needs to select how much of the transmit power is used for each service, which is represented by the variable $0 \leq \alpha \leq 1$, where $\alpha = 1$  and $\alpha = 0$ represent the extreme cases where the base station prioritizes communication only and radar only, respectively.
	
	Based on the above considerations, the power of communication and radar can be written as 
	\begin{eqnarray}
		\label{eq:P_c} P_{\rm C} &=& P(\beta + (1-\beta)\alpha), \label{eq:P_C} \\
		\text{and} \quad
		\label{eq:P_r}
		P_{\rm R} &=& P(\beta + (1-\beta)(1-\alpha)), \label{eq:P_R}
	\end{eqnarray}
	where $P$ is the transmit power. 
	We now define 
	\begin{eqnarray}
		\label{eq:P_c_loss} {L}_{{\rm C}_\beta} &=& (\beta + (1-\beta)\alpha),  \\
		\text{and} \quad
		\label{eq:P_r_loss}
		{L}_{{\rm R}_\beta} &=& (\beta + (1-\beta)(1-\alpha)), 
	\end{eqnarray}
	as the SNR losses with respect to the fully coupled channel where $\beta = 1$.
	In practice, $\alpha$ is chosen for a given $\beta$, then it is more convenient to treat $\beta$ as a fixed parameter, which is indicated by the subscript.	
	\subsection{Communication SNR}
	The communication SNR is defined as the SNR at the communication user, which is given by
	\begin{equation}\label{eq:rho_C}
		\rho_{\rm C} = \frac{P {L}_{{\rm C}_\beta} G_{\rm BS}G_{\rm UE}}{N_{\rm UE}B}\left(\frac{c}{f 4\pi} \right)^2\frac{1}{d_{\rm C}^2},
	\end{equation}
	where $G_{\rm BS}$ is the BS antenna gain, $G_{\rm UE}$ is the \ac{UE} antenna gain, $N_{\rm UE}$ is the UE noise \ac{PSD}, $B$ is the bandwidth, $c$ is the speed of light, $f$ is the center frequency and $d_{\rm C}$ is the distance between the BS and UE.
	\subsection{Radar SNR}
	The radar SNR is the echo SNR at the \ac{DFRC} base station, which is defined by
	\begin{equation}\label{eq:rho_R}
		\rho_{\rm R} = \frac{P {L}_{{\rm R}_\beta} G_{\rm BS}^2 G_{\rm P}}{N_{\rm BS}B}\left(\frac{c}{f}\right)^2\frac{\sigma_{\rm rcs}}{\left(4\pi\right)^3 d_{\rm R}^4},
	\end{equation}
	where $G_{\rm P}$ is the radar processing gain, $N_{\rm BS}$ is the \ac{PSD} of the noise at the BS, $\sigma_{\rm rcs}$ is the radar cross section and $d_{\rm R}$ is the distance between the BS and radar target. The term $G_{\rm BS}^2$ reflects the case where the transmit and receiver antennas at the BS have the same gain.
	
	\subsection{Possible extensions of the model}
	We briefly state possible approaches to the extension of the above model. 
	For the generalization to multi-user and multi-target, a proper power split among the extra nodes should be properly included in equations \eqref{eq:P_C} and \eqref{eq:P_R}.
	
	In addition, in order to include multi-path fading channel, random RCS of targets and clutter can be considered, where the SNRs in \eqref{eq:rho_C} and \eqref{eq:rho_R} should be seen as random variables.
	
	\section{SNR Operating Point (SOP) Equation}\label{sec:SNR_analysis}
	%
	%
	\subsection{SOP Derivation}
	In general, the positions of the communication user and target are different\footnote{The communication user and target can be in different locations as long as they are within the direction of the same transmit beam.}. 
	We can define the distance ratio which reflects how many times the communication user is farther than the target
	\begin{equation}\label{eq:delta}
		\delta = d_{\rm C}/d_{\rm R},
	\end{equation}
	which is typically greater than one since the radar SNR decays faster with distance.
	In order to match both SNRs in equations \eqref{eq:rho_C} and \eqref{eq:rho_R}, we swap $d_{\rm C}$ and $\rho_{\rm C}$ in \eqref{eq:rho_C}, and substitute the $d_{\rm R} = d_{\rm C}/\delta$ into \eqref{eq:rho_R} to have the SOP equation as
	\begin{equation}
		\begin{split}
			\rho_{\rm R} 
			= \frac{\rho_{\rm C}^2 {L}_{{\rm R}_\beta} }{P  {L}_{{\rm C}_\beta}^2 }\frac{ B  G_{\rm P} \sigma_{\rm rcs} \delta^4}{P G_{\rm UE}^2} \frac{N_{\rm UE}^2}{N_{\rm BS}}\frac{f^2 4 \pi}{c^2}.
		\end{split}
	\end{equation}
	%
	In the logarithm domain, $\tilde{\rho}_{\rm R} = 10\log {\rho}_{\rm R}$, we have	
	\begin{equation}\label{eq:snr_RC}
		\begin{split}
			\tilde{\rho}_{\rm R} =  & 2 \tilde{\rho}_{\rm C} + \tilde{L}_{{\rm R}_\beta} + \tilde{B}  + 2\tilde{f} + \tilde{G}_{\rm P} + \tilde{\sigma}_{\rm rcs} + 4\tilde{\delta} + \tilde{N}
			\\ &  -\tilde{P} -2\tilde{G}_{\rm UE}  - 2\tilde{L}_{{\rm C}_\beta} + \tilde{\nu},
		\end{split}
	\end{equation}
	where $\tilde{N} = 2\tilde{N}_{\rm UE} - \tilde{N}_{\rm BS}$ and $\tilde{\nu} = 10\log (4\pi /c^2)$ is a constant.
	\subsection{SOP Discussion}
	Basically, the SOP equation in \eqref{eq:snr_RC} defines a set of possible SNR pairs $(\tilde{\rho}_{\rm C},\tilde{\rho}_{\rm R})$ through a straight line for a given system parameter set, where each point has a direct and implicit correspondence to a distance pair $(d_{\rm C},d_{\rm R})$ with $d_{\rm R} = d_{\rm C}/\delta$ by substitution in equations \eqref{eq:rho_C} and \eqref{eq:rho_R}.
	In addition, \eqref{eq:snr_RC} is a linear equation with a slope of two, meaning that a 10 dB increase/decrease of $\tilde{\rho}_{\rm C}$ leads to a 20 dB increase/decrease of $\tilde{\rho}_{\rm R}$. 
	This happens because communication SNR decreases with the 2$nd$ power of distance, while the radar SNR decreases with the 4$th$ power of distance.
	\begin{figure}[t!]
		\centering
		\tikzstyle{block} = [draw, fill=white, rectangle, 
minimum height=3em, minimum width=3em]
\tikzstyle{block2} = [draw, fill=white, rectangle, 
minimum height=1em, minimum width=2.4em]

\tikzstyle{multiplier} = [draw,circle,fill=blue!20,add={}{}{}{}] {} 
\tikzstyle{sum} = [draw,circle,scale=0.7,add2={}{}{}{}] {} 
\tikzstyle{input} = [coordinate]
\tikzstyle{output} = [coordinate]
\tikzstyle{pinstyle} = [pin edge={to-,thin,black}]
\newcommand\z{3}

\def\windup{
	\tikz[remember picture,overlay]{
		\draw (-0.8,0) -- (0.8,0);
		\draw (-0.8,-0.4)--(-0.4,-0.4) -- (0.4,0.4) --(0.8,0.4);
	}}
	
	\centering
	\begin{tikzpicture}[auto, node distance=2cm,>=latex',scale=0.8]
	
%
%
%
%
%
%
%

	\draw[->] (-0.2, 0) -- (3, 0) node[below] {$\tilde{\rho}_{\rm C}$};
	\draw[->] (0, -0.2) -- (0, 4) node[left] {$\tilde{\rho}_{\rm R}$};
	\draw[scale=0.5, domain=-0.5:4.5, smooth, variable=\x, black] plot ({\x}, {2*\x}) node[right] {\footnotesize reference};
	\draw[scale=0.5, domain=-1:3.5, smooth, variable=\x, black] plot ({\x}, {2*\x+\z}) node[right] {\footnotesize increase of $\tilde{B},\tilde{f},\tilde{G}_{\rm P},\tilde{\sigma}_{\rm rcs},\tilde{\delta}$ or $\tilde{N}$};
	\draw[scale=0.5, domain=1:5.5, smooth, variable=\x, black] plot ({\x}, {2*\x-\z})node[right] {\footnotesize increase of $\tilde{P}$ or $\tilde{G}_{\rm UE}$};
	\draw node[xshift=0cm,yshift=-0.3cm] {\small $a$} node[xshift=-0.3cm,yshift=-0cm] {\small $b$};
	
	\draw[scale=0.5, domain=3:0, smooth, variable=\x, black,dashed] plot (3, {2*\x}) node[yshift=-0.15cm] {\footnotesize $a+x$};
	\draw[scale=0.5, domain=3:0, smooth, variable=\x, black,dashed] plot (\x, 6) node[xshift=-0.5cm,yshift=0cm] {\footnotesize $b+2x$};
	\draw plot[only marks,mark size = 2,mark={o}] coordinates{(1.5,3)} node[xshift=0.25cm] {\footnotesize $p_0$};
	
	\draw plot[only marks,mark size = 2,mark={*}] coordinates{(0,1.5)} node[xshift=0.25cm] {\footnotesize $p_1$};
	\draw plot[only marks,blue,fill opacity = 0.1,draw opacity=0, mark size = 2,mark={square*}] coordinates{(0.75,3)} node[xshift=0.1cm,yshift=-0.22cm] {\footnotesize $p_2$};
	\draw plot[only marks,mark size = 2,mark={*}] coordinates{(1.5,4.5)} node[xshift=0.25cm] {\footnotesize $p_3$};
	\draw plot[only marks,mark size = 2,mark={*}] coordinates{(2.5,3.5)} node[xshift=0.25cm] {\footnotesize $p_4$};
	\draw plot[only marks,mark size = 2,mark={*}] coordinates{(2.25,3)} node[xshift=0.25cm] {\footnotesize $p_5$};
	\draw plot[only marks,mark size = 2,mark={*}] coordinates{(2,4)} node[xshift=0.25cm] {\footnotesize $p_6$};

	\node[anchor=west] (text1) at (3.5,2.6) {\footnotesize $p_0 \rightarrow p_1$, increase of $\tilde{B}$, $\tilde{f}$ or $\tilde{N}$};
	\node[anchor=west] (text2) at ([yshift=-0.1cm]text1.south west) {\footnotesize $p_1 \rightarrow p_2$, decrease of $d_{\rm C}$};
	\node[anchor=west] (text3) at ([yshift=-0.1cm]text2.south west) {\footnotesize $p_0 \rightarrow p_3$, increase of $\tilde{G}_{\rm P}$, $\tilde{\sigma}_{\rm rcs}$ or $\tilde{\delta}$};
	\node[anchor=west] (text4) at ([yshift=-0.1cm]text3.south west) {\footnotesize $p_0 \rightarrow p_4$, increase of $\tilde{P}$};
	\node[anchor=west] (text5) at ([yshift=-0.1cm]text4.south west) {\footnotesize $p_0 \rightarrow p_5$, increase of $\tilde{G}_{\rm UE}$};
	\node[anchor=west] (text6) at ([yshift=-0.1cm]text5.south west) {\footnotesize $p_0 \rightarrow p_6$, increase of $\tilde{G}_{\rm BS}$};

	\end{tikzpicture}
	
		\vspace{-1cm}
		\caption{Qualitative illustration of equation \eqref{eq:snr_RC}. This graph shows the linear relation between $\tilde{\rho}_{\rm C}$ and $\tilde{\rho}_{\rm R}$ defined by each distance pair $(d_{\rm C},d_{\rm R})$ for a fixed ratio $\delta = d_{\rm C}/d_{\rm R}$. }
		\label{fig:snr}
		\vspace{-0.5cm}
	\end{figure}
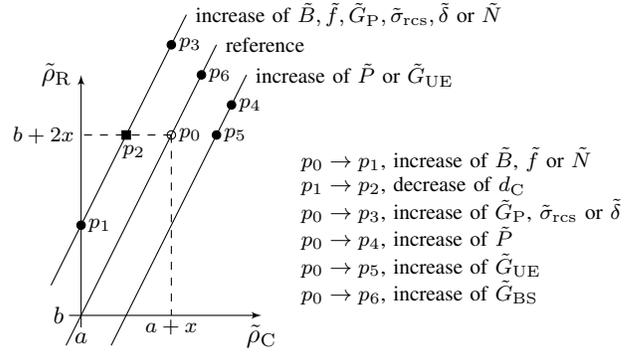
	%
	%
	The parameters in \eqref{eq:snr_RC} have the impact of shifting the straight line upwards or downwards.
	Moving this straight line upwards implies that a given $\tilde{\rho}_{\rm C}$ matches a larger $\tilde{\rho}_{\rm R}$, which is favorable to the radar application.
	An opposite effect happens by shifting the linear function downwards. 
	
	In the following, we discuss the role of each parameter in the SNR equation excluding $\tilde{L}_{{\rm C}_\beta}$ and $\tilde{L}_{{\rm R}_\beta}$, which are treated in detail in the Section~\ref{sec:analaysis_alpha_beta}.
	By analyzing \eqref{eq:snr_RC}, we are able to identify that increasing $\tilde{B},\tilde{f},\tilde{G}_{\rm P},\tilde{\sigma}_{\rm rcs},\tilde{\delta}$ or $\tilde{N}$ shifts the line upwards leading to an SOP that favors radar, whereas increasing $\tilde{P}$ or $\tilde{G}_{\rm UE}$ shift it downwards leading to an SOP that favors communication.
	It is also noticeable that $\tilde{G}_{\rm BS}$ disappears in \eqref{eq:snr_RC}, although it has an impact in the SNR equations \eqref{eq:rho_C} and \eqref{eq:rho_R}.
	It means that changing $\tilde{G}_{\rm BS}$ does not modify the allowed SNR pairs.
	These observations are summarized in the three statements below.
	
	\begin{statement}
		Increasing $\tilde{B},\tilde{f},\tilde{G}_{\rm P},\tilde{\sigma}_{\rm rcs},\tilde{\delta}$ or $\tilde{N}$ allows an SNR operation point with higher radar SNR for the same communication SNR. Decreasing these parameters has the opposite effect.
	\end{statement}
	
	\begin{statement}
		Increasing $\tilde{P}$ or $\tilde{G}_{\rm UE}$ allows an SNR operation point with lower radar SNR for the same communication SNR. Decreasing these parameters has the opposite effect.
	\end{statement}
	
	\begin{statement}
		Increasing or decreasing $\tilde{G}_{\rm BS}$ does not change the set of possible SNR pairs for radar and communication.
	\end{statement}
	
	Figure \ref{fig:snr} provides an illustration of equation \eqref{eq:snr_RC} and Statements 1-3.
	A reference curve for an arbitrary parameter set is drawn passing through the points $(a,b)$ and $(a+x,b+2x)$.
	The point $p_0$ represents an arbitrary pair of SNRs $(\tilde{\rho}_{{\rm C}_0},\tilde{\rho}_{{\rm R}_0})$ for a given distance pair $(d_{{\rm C}_0},d_{{\rm C}_0}/\delta)$.
	The point $p_1$ lies in the upwards shifted line for an increased value of $\tilde{B}$, $\tilde{f}$, or $\tilde{N}$ in relation to the reference line.
	Notice that increasing $\tilde{B}$, $\tilde{f}$ or $\tilde{N}$ decreases the SNRs of both communication and radar equally in \eqref{eq:rho_C} and \eqref{eq:rho_R}.
	Taking $p_1$ as reference, we can define the point $p_2$ by decreasing $d_{\rm C}$, where $\tilde{\rho}_{{\rm C}_2} < \tilde{\rho}_{{\rm C}_0}$ while $\tilde{\rho}_{{\rm R}_2}=\tilde{\rho}_{{\rm R}_0}$. 
	This is equivalent to Statement 1, since increasing $\tilde{B}$, $\tilde{f}$, or $\tilde{N}$ also allows an operation point with higher radar SNR for the same communication SNR.
	The point $p_3$ is reached for increased values of $\tilde{G}_{\rm P}$, $\tilde{\sigma}_{\rm rcs}$ or $\tilde{\delta}$ in relation to the $p_0$.
	The points $p_4$ and $p_5$ lie in the downwards shifted line for an increased value of $\tilde{P}$ and $\tilde{G}_{\rm UE}$ in relation to the reference line, respectively.
	Using an equivalent argument as before, we can state that increasing $\tilde{P}$ or $\tilde{G}_{\rm UE}$ allows an operation point with lower radar SNR for the same communication SNR, which is equivalent to Statement 2.
	As shown in Figure \ref{fig:snr}, the point $p_6$ has increased $\tilde{G}_{\rm BS}$ in relation to $p_0$ and lies in the reference line, illustrating Statement 3.

	\section{Analysis of $\alpha$ and $\beta$ in the SNR Equation}\label{sec:analaysis_alpha_beta}
	%
	The trade-off and coupling parameters $\alpha$ and $\beta$ scale the communication and radar SNRs according to the terms ${L}_{{\rm C}_\beta} = \beta + (1-\beta)\alpha$ and ${L}_{{\rm R}_\beta} = \beta + (1-\beta)(1-\alpha)$ defined in \eqref{eq:P_c_loss} and \eqref{eq:P_r_loss}, respectively, which are interpreted as SNR losses with respect to the fully coupled channel where $\beta = 1$.
	For a fixed $\beta$, we relate the SNR loss pair $({L}_{{\rm R}_\beta},{L}_{{\rm C}_\beta})$ as
	\begin{equation}\label{eq:SNR_RC_beta}
		{L}_{{\rm R}_\beta} = 1+\beta - {L}_{{\rm C}_\beta},
	\end{equation}
	for $\beta \leq {L}_{{\rm C}_\beta},{L}_{{\rm R}_\beta} \leq 1$, where each pair $({L}_{{\rm R}_\beta},{L}_{{\rm C}_\beta})$ has a direct correspondence to a value of $0\leq \alpha \leq 1$. 
	In particular, the extreme cases of $(P_{{\rm R}_\beta}' = \beta, L_{{\rm C}_\beta} = 1)$ and $(L_{{\rm R}_\beta} = 1, L_{{\rm C}_\beta} = \beta)$ are obtained for $\alpha = 1$ and $\alpha = 0$, respectively.
	Notice that for $\beta = 1$, the value of $\alpha$ is irrelevant because the channel is fully coupled and there is no power trade-off between communication and radar.
	The counterpart of \eqref{eq:SNR_RC_beta} in the log domain is
	\begin{equation}\label{eq:SNR_RC_beta_log}
		\tilde{L}_{{\rm R}_\beta} = 10\log\left(1+\beta - 10^{\tilde{L}_{{\rm C}_\beta}/10}\right),
	\end{equation}
	for $10\log\beta \leq \tilde{L}_{{\rm C}_\beta},\tilde{L}_{{\rm R}_\beta} \leq 0$.	
	Equations \eqref{eq:SNR_RC_beta} and \eqref{eq:SNR_RC_beta_log} are depicted in Fig.~\ref{fig:snr_generalized} for different channel coupling levels $\beta$.
	In the logarithmic scale (right graph), we also plot equation \eqref{eq:snr_RC} as a reference to establish a relation between the analysis done in Section~\ref{sec:SNR_analysis}\footnote{A concrete example is given in Section \ref{sec:numerical_examples}, Fig.~\ref{fig:snr_analysis}.}.
	Basically, the SNR loss pair $(\tilde{L}_{{\rm C}_\beta},\tilde{L}_{{\rm R}_\beta}) = (0,0)$ corresponds to the SNR pair $(\tilde{\rho}_{{\rm C}},\tilde{\rho}_{{\rm R}})$ in \eqref{eq:snr_RC} for the case of $\beta=1$.
	And since the SNR loss pair $(\tilde{L}_{{\rm C}_\beta},\tilde{L}_{{\rm R}_\beta})$ for any $\beta$ and $\alpha$ combination can always be found in relation to $(\tilde{L}_{{\rm C}_\beta},\tilde{L}_{{\rm R}_\beta}) = (0,0)$, we establish without loss of generality that i) \eqref{eq:snr_RC} can assume $\beta=1$ defining $(\tilde{\rho}_{{\rm C}},\tilde{\rho}_{{\rm R}})_{\beta=1}$, and ii) the resulting SNR for any $\beta$ and $\alpha$ combination is found by subtracting the SNR loss pair $(\tilde{L}_{{\rm C}_\beta},\tilde{L}_{{\rm R}_\beta})$ from the SNR pair $(\tilde{\rho}_{{\rm C}},\tilde{\rho}_{{\rm R}})_{\beta=1}$.
	
	
	%
	\section{ISAC Range Equation}\label{sec:range_equation}
	%
	An analogous analysis of the ISAC SNR can be carried out for the range of both systems.
	In some applications such as sensing-assisted beamforming, the range for radar and communication should be in the same order of magnitude for efficient power usage.
	In this case, we are interested in evaluating the ratio $\delta$ in \eqref{eq:delta} and how the system parameters impact it.
	For instance, if $\delta=100$, it implies that the range for communications is 100 times larger than the range for radar, meaning that the communication link will operate in an SNR region much higher than necessary, which is inefficient.
	
	In order to analyze $\delta$, we isolate $\tilde{\delta}$ in \eqref{eq:snr_RC} as
	\begin{equation}\label{eq:delta_db}
			\tilde{\delta}  = \frac{\tilde{\rho}_{\rm R}^\star}{4} - \frac{\tilde{\rho}_{\rm C}^\star}{2} - \frac{\tilde{B}}{4}  - \frac{\tilde{f}}{2} - \frac{\tilde{G}_{\rm P}}{4}-\frac{\tilde{\sigma}_{\rm rcs}}{4} - \frac{\tilde{N}}{4} + \frac{\tilde{P}}{4}+ \frac{\tilde{G}_{\rm UE}}{2} - \frac{\tilde{\nu}}{4} + {\tilde{\delta}_\beta}
	\end{equation}
	where $\tilde{\rho}_{\rm R}^\star$ and $\tilde{\rho}_{\rm C}^\star$ represent the desired SNR for radar and communication, respectively, and
	\begin{equation}\label{eq:delta_beta}
		{\tilde{\delta}_\beta} = \frac{10}{4}\log \left(\frac{(\beta + (1-\beta)\alpha)^2}{\beta + (1-\beta)(1-\alpha)} \right)
	\end{equation}
	is found by substituting the quantities defined in \eqref{eq:P_c_loss} and \eqref{eq:P_r_loss} and takes into account the impact of $\alpha$ and $\beta$.
	Following an analogous analysis as in the last subsection, an equivalent set of statements regarding the ratio between the communication and radar distances $\delta =d_{\rm C}/d_{\rm R}$ for a fixed pair of desired SNRs $\tilde{\rho}_{\rm R}^\star$ and $\tilde{\rho}_{\rm C}^\star$ can be expressed. 
	Particularly, increasing $\tilde{B},\tilde{f},\tilde{G}_{\rm P},\tilde{\sigma}_{\rm rcs}$ or $\tilde{N}$ decreases $\delta$.
	Increasing $\tilde{P}$ or $\tilde{G}_{\rm UE}$ increases the $\delta$, and $\tilde{G}_{\rm BS}$ does not impact $\delta$.

	\setcounter{figure}{2}
	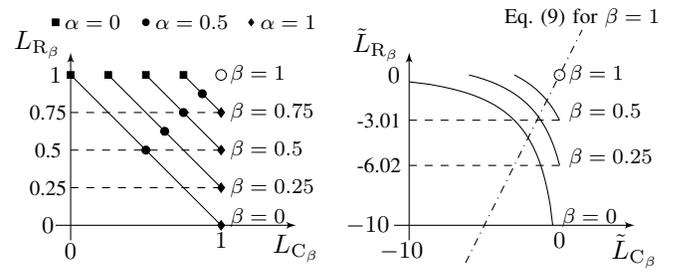
\begin{figure}[t!]
		\centering
		\tikzstyle{block} = [draw, fill=white, rectangle, 
minimum height=3em, minimum width=3em]
\tikzstyle{block2} = [draw, fill=white, rectangle, 
minimum height=1em, minimum width=2.4em]

\tikzstyle{multiplier} = [draw,circle,fill=blue!20,add={}{}{}{}] {} 
\tikzstyle{sum} = [draw,circle,scale=0.7,add2={}{}{}{}] {} 
\tikzstyle{input} = [coordinate]
\tikzstyle{output} = [coordinate]
\tikzstyle{pinstyle} = [pin edge={to-,thin,black}]
\newcommand\z{4.5}

\def\windup{
	\tikz[remember picture,overlay]{
		\draw (-0.8,0) -- (0.8,0);
		\draw (-0.8,-0.4)--(-0.4,-0.4) -- (0.4,0.4) --(0.8,0.4);
}}

\centering
\begin{tikzpicture}[auto, node distance=0cm,>=latex']

	\draw[->] (-0.2, 0) -- (3, 0) node[below] {$L_{{\rm C}_\beta}$};
	\draw[->] (0, -0.2) -- (0, 2.3) node[left,yshift = 0.1cm] {$L_{{\rm R}_\beta}$};	
	
	\draw[scale=2, domain=0:1, smooth, variable=\x, black] plot ({\x}, {1-\x}) node[right,xshift = -0cm,yshift = 0.1cm] {\footnotesize $\beta = 0$}; 
	\draw[scale=2, domain=0.25:1, smooth, variable=\x, black] plot ({\x}, {1+0.25-\x}) node[right,xshift = -0cm,yshift = 0cm] {\footnotesize $\beta = 0.25$}; 
	\draw[scale=2, domain=0.5:1, smooth, variable=\x, black] plot ({\x}, {1+0.5-\x}) node[right,xshift = -0cm,yshift = 0cm] {\footnotesize $\beta = 0.5$}; 
	\draw[scale=2, domain=0.75:1, smooth, variable=\x,black] plot ({\x}, {1+0.75-\x}) node[right,xshift = -0cm,yshift = 0cm] {\footnotesize $\beta = 0.75$}; 
	\draw plot[only marks,mark size = 2,mark={o}] coordinates{(2,2)} node[right,xshift = -0cm,yshift = 0cm] {\footnotesize $\beta = 1$};	
	
	\draw plot[only marks,mark size = 1.25,mark={square*}] coordinates{(0,2)} node[right,xshift = -0cm,yshift = 0cm] {};	
	\draw plot[only marks,mark size = 1.25,mark={square*}] coordinates{(0.5,2)} node[right,xshift = -0cm,yshift = 0cm] {};	
	\draw plot[only marks,mark size = 1.25,mark={square*}] coordinates{(1,2)} node[right,xshift = -0cm,yshift = 0cm] {};	
	\draw plot[only marks,mark size = 1.25,mark={square*}] coordinates{(1.5,2)} node[right,xshift = -0cm,yshift = 0cm] {};			
	
	\draw plot[only marks,mark size = 1.5,mark={*}] coordinates{(1,1)} node[right,xshift = -0cm,yshift = 0cm] {};	
	\draw plot[only marks,mark size = 1.5,mark={*}] coordinates{(1.25,1.25)} node[right,xshift = -0cm,yshift = 0cm] {};	
	\draw plot[only marks,mark size = 1.5,mark={*}] coordinates{(1.5,1.5)} node[right,xshift = -0cm,yshift = 0cm] {};	
	\draw plot[only marks,mark size = 1.5,mark={*}] coordinates{(1.75,1.75)} node[right,xshift = -0cm,yshift = 0cm] {};	
	
	\draw plot[only marks,mark size = 1.75,mark={diamond*}] coordinates{(2,0)} node[right,xshift = -0cm,yshift = 0cm] {};	
	\draw plot[only marks,mark size = 1.75,mark={diamond*}] coordinates{(2,0.5)} node[right,xshift = -0cm,yshift = 0cm] {};	
	\draw plot[only marks,mark size = 1.75,mark={diamond*}] coordinates{(2,1)} node[right,xshift = -0cm,yshift = 0cm] {};	
	\draw plot[only marks,mark size = 1.75,mark={diamond*}] coordinates{(2,1.5)} node[right,xshift = -0cm,yshift = 0cm] {};	
	
	\draw plot[only marks,mark size = 1.25,mark={square*}] coordinates{(-0.2,2.7)} node[right,xshift = -0cm,yshift = 0cm] {\footnotesize $\alpha = 0$};	
	\draw plot[only marks,mark size = 1.25,mark={*}] coordinates{(-0.2+1.2,2.7)} node[right,xshift = -0cm,yshift = 0cm] {\footnotesize $\alpha = 0.5$};	
	\draw plot[only marks,mark size = 1.25,mark={diamond*}] coordinates{(-0.2+2.6,2.7)} node[right,xshift = -0cm,yshift = 0cm] {\footnotesize $\alpha = 1$};	
			
	\draw node[xshift=0cm,yshift=-0.35cm] {\small $0$} node[xshift=-0.3cm,yshift=-0cm] {\small $0$};
	\draw node[xshift=2cm,yshift=-0.2cm] {\small $1$} node[xshift=-0.2cm,yshift=2cm] {\small $1$};
	
	\draw[scale=2, domain=1:0, smooth, variable=\x, black,dashed] plot ({\x},0.75) node[xshift=-0.3cm] {\footnotesize 0.75};
	\draw[scale=2, domain=1:0, smooth, variable=\x, black,dashed] plot ({\x},0.5) node[xshift=-0.3cm] {\footnotesize 0.5};
	\draw[scale=2, domain=1:0, smooth, variable=\x, black,dashed] plot ({\x},0.25) node[xshift=-0.3cm] {\footnotesize 0.25};

	\draw[->] (\z-0.2, 0) -- (3+\z, 0) node[below] {$\tilde{L}_{{\rm C}_\beta}$};
	\draw[->] (0+\z, -0.2) -- (0+\z, 2.3) node[left,yshift = 0.1cm] {$\tilde{L}_{{\rm R}_\beta}$};	
	\draw[shift = {(\z+2,2)},scale=2, domain=0.1:0.9, variable=\x, black,samples=25] plot ({log10{\x}}, {log10{(1-\x)}}) node[right,xshift = -0cm,yshift = 0.1cm] {\footnotesize $\beta = 0$}; 
	\draw[shift = {(\z+2,2)},scale=2, domain=0.25:0.999, variable=\x, black,samples=25] plot ({log10{\x}}, {log10{(1.25-\x)}}) node[right,xshift = -0cm,yshift = 0.1cm] {\footnotesize $\beta = 0.25$}; 
	\draw[shift = {(\z+2,2)},scale=2, domain=0.5:0.999, variable=\x, black,samples=25] plot ({log10{\x}}, {log10{(1.5-\x)}}) node[right,xshift = -0cm,yshift = 0.1cm] {\footnotesize $\beta = 0.5$}; 
	\draw [shift = {(\z+2,2)}] plot[only marks,mark size = 2,mark={o}] coordinates{(0,0)} node[right,xshift = -0cm,yshift = 0cm] {\footnotesize $\beta = 1$};	
	
	\draw [shift = {(\z,0)}] node[xshift=-0.1cm,yshift=-0.35cm] {\small $-10$} node[xshift=-0.55cm,yshift=-0cm] {\small $-10$};
	\draw [shift = {(\z,0)}] node[xshift=2cm,yshift=-0.2cm] {\small $0$} node[xshift=-0.2cm,yshift=2cm] {\small $0$};
	\draw[shift = {(\z,0)},scale=2, domain=1:0, smooth, variable=\x, black,dashed] plot ({\x},1-0.301) node[xshift=-0.4cm] {\footnotesize -3.01};
	\draw[shift = {(\z,0)},scale=2, domain=1:0, smooth, variable=\x, black,dashed] plot ({\x},1-0.602) node[xshift=-0.4cm] {\footnotesize -6.02};
		
	\draw[shift = {(\z+1,0)},scale=1, domain=-0.25:1.3, variable=\x, black,samples=25,dash dot] plot (\x, 2*\x) node[above,xshift = -0.0cm,yshift = -0.1cm] {\footnotesize Eq. \eqref{eq:snr_RC} for $\beta=1$}; 
\end{tikzpicture}
		\vspace{-1cm}
		\caption{Illustration of equations \eqref{eq:SNR_RC_beta} and \eqref{eq:SNR_RC_beta_log} in the linear (left) and logarithmic (right) scales, respectively, for different coupling levels $\beta$. For a fixed $\beta$, a pair $(\rho_{{\rm C}_\beta}',\rho_{{\rm R}_\beta}')$ is defined for each $\alpha$.}
		\label{fig:snr_generalized}
		\vspace{-0.5cm}
	\end{figure}
	
	\setcounter{figure}{3}
	\begin{figure*}
		\centering
		\input{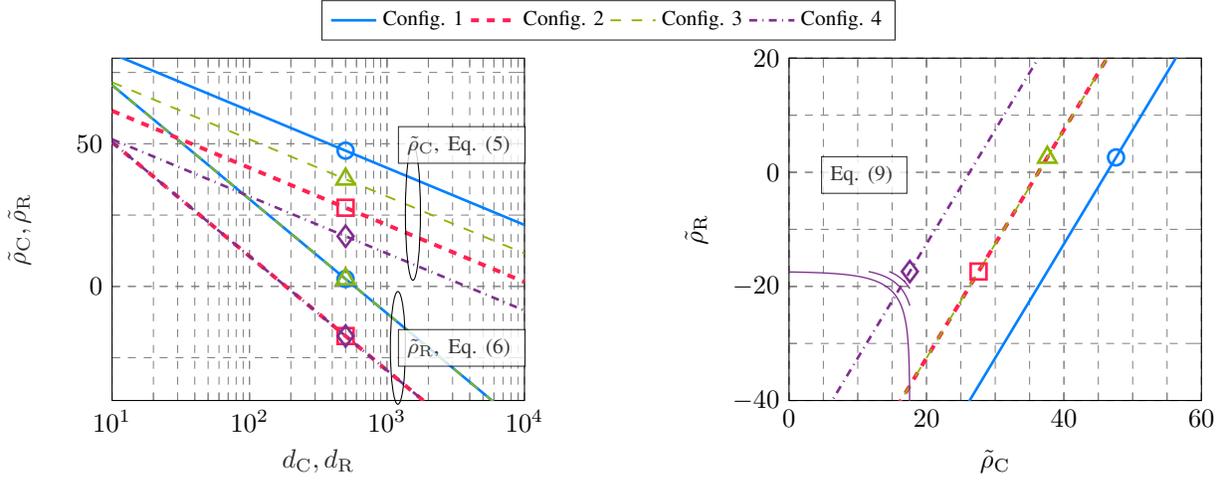} 
		\vspace{-0.3cm}
		\caption{SNR Operating Point analysis for the configurations of Table~\ref{tab:snr_analysis}$^1$. Left side: communication and radar SNRs from \eqref{eq:rho_C} and \eqref{eq:rho_R} with reference points for $d_{\rm C}=d_{\rm R}=\SI{500}\m$. Right side: ISAC SNR equation from \eqref{eq:snr_RC}.}
		\label{fig:snr_analysis}
		\vspace{-0.5cm}
	\end{figure*}

	\section{Numerical Examples}\label{sec:numerical_examples}
		\begin{table}[t!]
		\centering
		\caption{Parameters for SNR analysis.}
		\vspace{-0.2cm}
		\label{tab:snr_analysis}
		\begin{tabular}{lcccc}
			\toprule
			Param. & {Config. 1} & {Config. 2} & {Config. 3} & {Config. 4} \\
			\midrule
			$f$					&$\SI{1} \GHz$ & $\SI{10} \GHz$ & $\SI{10} \GHz$ & $\SI{100} \GHz$ \\
			$B$						&$\SI{100} \MHz$ & $\SI{100} \MHz$ & $\SI{100} \MHz$ & $\SI{1} \GHz$\\
			$G_{\rm BS}$			&$10$ & $10$ & $100$ & $1000$ \\
			$\sigma_{\rm RCS}$		&$\SI{10}\m^2$ & $\SI{10}\m^2$ & $\SI{10}\m^2$ & $\SI{1}\m^2$ \\
			\bottomrule
		\end{tabular}
		\vspace{-0.2cm}
	\end{table}
	
	\subsection{SNR Operating Point Analysis}
	In order to provide a numerical example to illustrate statements 1-3 of Section~\ref{sec:SNR_analysis},  we consider four configurations for the SNR analysis given in Table~\ref{tab:snr_analysis}\footnote{Proper unit transformation should be done to apply these variables to equations \eqref{eq:rho_C}, \eqref{eq:rho_R} and \eqref{eq:snr_RC}.}.
	We highlight that from configuration 1 to 2, the center frequency $f$ is increased 10 times.
	From configurations 2 to 3, the BS antenna gain $G_{\rm BS}$ is increased 10 times.
	From configuration 3 to 4, the center frequency, BS antenna gain, and bandwidth $B$ are increased 10 times, while the RCS is decreased 10 times since we expect to detect smaller objects as the frequency increases.
	The values $G_{\rm UE}=4$, ${N}_{\rm BS}={N}_{\rm UE}=-174 \,{\rm dBm}$ are the same for all configurations.
	Also, this analysis ignores the radar processing gain by setting $\tilde{G}_{\rm P} = 0$.
	Lastly, equations \eqref{eq:rho_C}, \eqref{eq:rho_R} and \eqref{eq:snr_RC} consider $\beta=1$ as discussed in Section~\ref{sec:analaysis_alpha_beta}.
	
	The results are reported in Fig.~\ref{fig:snr_analysis}.
	We first observe that from configuration 1 to 2, there is a loss of 20 dB for both communication and radar SNR due to a 10 times increase in center frequency.
	Configuration 3 increases $G_{\rm BS}$, since the radar SNR in \eqref{eq:rho_R} is proportional to $G_{\rm BS}^2$ and the communication SNR in \eqref{eq:rho_C} is proportional to $G_{\rm BS}$, the increase in decibels corresponds to 20 and 10, respectively.
	By observing the right side graph, lines corresponding to configurations 2 and 3 are above the line of configuration 1, as predicted by \eqref{eq:snr_RC} as the frequency increases.
	Also, for the same $d_{\rm C} = d_{\rm R} = \SI{500}\m$, the SNR pair for configuration 3 is at a higher point than configuration 2.
	Comparing configurations 1 and 3, it is clear that the communication SNR has dropped while the radar SNR has remained constant, which is an illustration of Statement 1 in Section~\ref{sec:SNR_analysis}.
	Another interesting example is comparing configurations 2 and 4.
	The left side graph shows that configuration 2 provides higher SNR for communication while configurations 3 and 4 overlap for radar.
	Basically, from configuration 2 to 4, BS antenna gain has increased 100 times, which compensates for the SNR loss of radar due to higher frequency and bandwidth. 
	For communication, however, the antenna gain increment at the BS is not sufficient to combat the same SNR loss.
	For completeness, the SNR pairs for $\beta = \left(0,0.25,0.5\right)$ and varying $0<\alpha<1$ are shown for Configuration 4 taking the SNR point $(\tilde{\rho}_{{\rm C}},\tilde{\rho}_{{\rm R}})_{\beta=1} = (17.5 {\rm dB}, -17.4 {\rm dB})$ as reference, in analogy to Fig.~\ref{fig:snr_generalized}.
	
	These outcomes reveal that there is a trade-off in favor of radar as the ISAC system moves towards higher frequencies, where the small wavelength allows a relatively larger number of antennas at the monostatic radar unit.

	\subsection{Range Analysis}
	For the range analysis, the goal is to compare the ratio $\delta = d_{\rm C}/d_{\rm R}$ in equation \eqref{eq:delta_db} for a given desired communication and radar SNR pair $(\tilde{\rho}_{\rm C}^\star,\tilde{\rho}_{\rm R}^\star)$.
	One useful analysis is to compute $\delta$ depending on the spectral efficiency $R = \log_2(1+\rho_{\rm C})$ in bits/s/Hz for a fixed value of $\tilde{\rho}_{\rm R}^\star$ assuming Gaussian signaling.
	In this case, the communication SNR is replaced by the spectrum efficiency with the transformation $\tilde{\rho}_{\rm C}^{\star} = 10\log (2^R-1)$.
	We set the radar SNR as \mbox{$\tilde{\rho}_{\rm R}^\star = 10.8$ dB} which is sufficient to attain a probability of detection of 0.9 for a constant probability of false alarm equal to $10^{-3}$, where coherent energy detection with the integration of 1024 samples corresponding to a processing gain of $\tilde{G}_{\rm P} = 30.1$ dB.
	Moreover, we consider $\beta=1$ for this analysis, which makes $\tilde{\delta}_\beta$ in \eqref{eq:delta_beta} equal to zero.
	For this evaluation, we define a new set of parameters in Table~\ref{tab:range_analysis}\footnote{Proper unit transformation should be done to apply these variables to equation \eqref{eq:snr_RC}.}, which are denoted sub-$\SI{6}\GHz$, mmWave and sub-$\SI{}\THz$ based on $f$.
	Table~\ref{tab:distance} shows for which distance $d_{\rm R}$ the radar SNR $\tilde{\rho}_{\rm R}^\star = 10.8$ dB is achieved. 
	
	\setcounter{figure}{4}
	\begin{figure}[t!]
		\centering
		\input{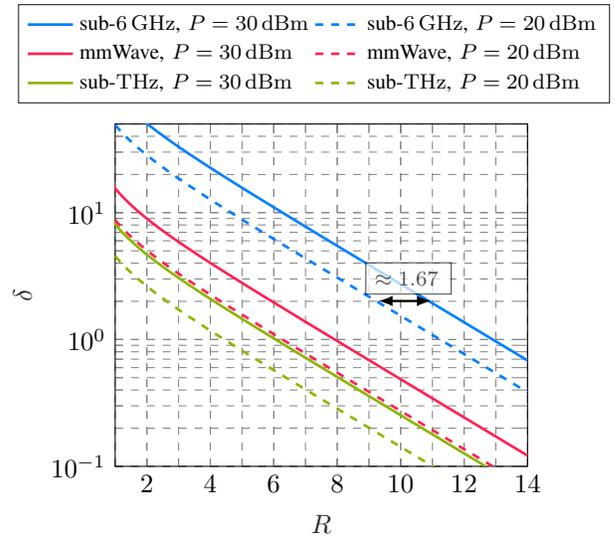} 
		\vspace{-0.4cm}
		\caption{Range ratio analysis for the configurations of Table.~\ref{tab:range_analysis}.}
		\label{fig:range_analysis}
		\vspace{-0.3cm}
	\end{figure}
	
	\begin{table}[t!]
		\centering
		\caption{Parameters for Range analysis.}
		\vspace{-0.2cm}
		\label{tab:range_analysis}
		\begin{tabular}{lccc}
			\toprule
			Param.		& (sub-$\SI{6}\GHz$) & (mmWave) & (sub-$\SI{}\THz$)  \\
			\midrule
			$f$	& $\SI{2.4} \GHz$ & $\SI{24} \GHz$ & $\SI{140} \GHz$ \\
			$B$	& $\SI{100} \MHz$ & $\SI{1} \GHz$ & $\SI{4} \GHz$ \\
			$G_{\rm BS}$	& $16$ & $64$ & $128$  \\
			$\sigma_{\rm RCS}$		&$\SI{10}\m^2$ & $\SI{10}\m^2$  & $\SI{1}\m^2$ \\
			\bottomrule
		\end{tabular}
	\end{table}
	\begin{table}[t!]
		\centering
		\caption{Target distance $d_{\rm R}$ for which $\rho_{\rm R}^\star = 10.8\,{\rm dB}$.}
		\vspace{-0.2cm}
		\label{tab:distance}
		\begin{tabular}{lccc}
			\toprule
			Tx Power	& (sub-$\SI{6}\GHz$) & (mmWave) & (sub-$\SI{}\THz$)  \\
			\midrule
			$P = 30 \, {\rm dBm}$	& $\SI{1442} \m$   & $\SI{513.4} \m$  & $\SI{119.4} \m$  \\
			$P = 20 \, {\rm dBm}$	& $\SI{811.5} \m$ & $\SI{288.6} \m$  & $\SI{67.1} \m$    \\
			\bottomrule
		\end{tabular}
	\end{table}
	
	The results are reported in Fig.~\ref{fig:range_analysis}.
	Special attention is given to the points crossing $\delta=10^0$, which reflects the situation $d_{\rm C}=d_{\rm R}$ that is relevant for sensing-assisted beamforming applications.
	For the sub-$\SI{6}\GHz$ systems with $P=30\,{\rm dBm}$, $\delta=10^0$ happens for $R\approx 12.9$ which is an extremely high spectral efficiency.
	For $P=20\,{\rm dBm}$, $R$ is reduced to 11.2, which is still very high.
	Since the communication typically operates within a lower spectral efficiency range, this implies that the distance between the BS and the communication can be considerably higher than the distance between the BS and the target in general. 
	A concrete example is given in the following to better illustrate the point. 
	From Table~\ref{tab:distance}, we see that for the sub-$\SI{6}{\GHz}$ with $P = 30 \, {\rm dBm}$, the radar target with distance $d_{\rm R} = \SI{1442}{\m}$ can be detected. 
	For communication, we consider a spectral efficiency of 8 bit/s/Hz as a reasonable practical value, which matches the value $\delta \approx 5.4$ in Fig.~\ref{fig:range_analysis}, from which we can extract  $d_{\rm C} = \delta d_{\rm R}  \approx \SI{7786}{\m}$.
	In this case, the discrepancy between the target and user distances is clear.
	Considering now the mmWave configuration again with $P = 30 \, {\rm dBm}$ with $R = 8$ bit/s/Hz, $\delta \approx 1$, meaning that $d_{\rm C} \approx d_{\rm R} = \SI{513.4}{\m}$ using the data from Table~\ref{tab:distance}.
	
	\begin{table}[t!]
		\centering
		\caption{Ratio $\delta$ and communication user distance $d_{\rm C}$ for which $R=2$ bits/s/Hz for $P=20 \, {\rm dBm}$.}
		\vspace{-0.2cm}
		\label{tab:distance_communication}
		\begin{tabular}{lccc}
			\toprule
			& (sub-$\SI{6}\GHz$) & (mmWave) & (sub-$\SI{}\THz$)  \\
			\midrule
			$\delta$	& 30   & 5  & 2.7  \\
			$d_{\rm C}$	& $\SI{24345} \m$ & $\SI{1443} \m$  & $\SI{181.2} \m$    \\
			\bottomrule
		\end{tabular}
		\vspace{-0.3cm}
	\end{table}
	
	In general, decreasing the transmit power shifts $\delta$ downwards, which is clear from \eqref{eq:delta_db}.
	This can be intuitively explained by the fact that the slope in which the radar SNR decreases with distance is steeper, meaning that the radar SNR increases at a higher rate than the communication SNR as the distance decreases, resulting in a smaller $\delta$.
	This is seen in Figure \ref{fig:range_analysis} since the curve with $P=20\,{\rm dBm}$ is below the curve with $P=30\,{\rm dBm}$.
	Moreover, we found that for each $10\,{\rm dB}$ of decreased transmit power, the corresponding data rate decrease is approximately $1.67$ bits/s/Hz, which is a useful reference number for the design of ISAC systems
	
	
	Another possible analysis is to analyze $\delta$ considering a minimum data rate for communications of $R=2$ bits/s/Hz, for the radar SNR $\tilde{\rho}_{\rm R}^\star = 10.8$.
	The corresponding values for $\delta$ and $d_{\rm C}$ are shown in Table~\ref{tab:distance_communication}.
	The ratio $\delta$ is found as 30, 5 and 2.7 for sub-$\SI{6}\GHz$, mmWave and sub-$\SI{}\THz$, respectively, which decreases with $f$ as expected. The distances $d_{\rm C}$ found as $\SI{24345} \m$, $\SI{1443} \m$ and $\SI{181.2} \m$ sub-$\SI{6}\GHz$, mmWave and sub-$\SI{}\THz$, respectively.
	
	In summary, the outcomes of this section reveal that ISAC systems tend to have a more balanced use of resources with higher frequencies such as mmWave and sub-$\SI{}\THz$.
	Regarding the mmWave and sub-$\SI{}\THz$ scenarios for $P=30\,{\rm dBm}$, $\delta=10^0$ is crossed for $R\approx 8$ and $R\approx 6$, which are values that meet practical transceiver implementations for the communication service. 
	Lastly, we note that we did not consider the impact of the trade-off $\alpha$ when the channel is not fully coupled due to the lack of space.
	However, such analysis can be conducted based \eqref{eq:delta_beta} which represents a shift of $\delta$ in the log scale. 
	Possibly, the discrepancy between the range for communications and radar can be controlled by $\alpha$ when $\beta<1$.
	
	\section{Conclusion}\label{sec:conclusion}
	In this paper, we have derived the SNR and range expressions of ISAC systems, which enable a system level analysis of single-target and single-user ISAC scenarios.
	For the SNR analysis, we have shown that with the increase of frequency and bandwidth, the ISAC SNR equation favors the radar application since for the same radar SNR, the communication SNR is decreased.
	The transmit power and antenna gain at the UE has the opposite effect.
	We have applied the range analysis to sensing-assisted beamforming applications, where the radar target and communication user are in the same location. 
	In particular, we have shown that for sub-$\SI{6}\GHz$ systems, the communication SNR is so high that implies impractically high spectral efficiency, while mmWave and sub-$\SI{}\THz$ scenarios lead to feasible rates, resulting in a practically acceptable radar and communication performance. 
	In summary, if the range for communication and radar is desired to be in the same order, the ISAC system should operate in mmWave or sub-$\SI{}\THz$ bands, while sub-$\SI{6}\GHz$ allows scenarios where the communication range is of orders of magnitude higher than radar.
	
		For future work, we propose the generalization to multi-user and multi-target, where a power split among the nodes should be properly included in the model.
		In addition, multi-path fading channels and random RCS of targets is of interest, where the SNRs should be seen as random variables.
	
	%
	%
	%
	%
	%
	%
	
	\bibliographystyle{ieeetr}
	\bibliography{references_ha}{}

\end{document}